\documentclass[preprint,number,hidelinks]{elsarticle}
\usepackage{amssymb}
\usepackage{amsmath}
\usepackage{bm}
\usepackage[dvipsnames,table]{xcolor}
\usepackage{hyperref}
\usepackage{verbatim}

\begin{document}

\begin{frontmatter}

\title{Optimal boron-doped graphene substrate for glucose Raman signal enhancement}

\author[1]{Jan Komeda \fnref{fn1}}
\ead{komedjan@fel.cvut.cz}
\author[1]{Antonio Cammarata \fnref{fn2}}
\ead{cammaant@fel.cvut.cz}
\author[1,2]{Tomas Polcar}
\affiliation[1]{organization={Department of Control Engineering, Faculty of Electrical Engineering,
Czech Technical University in Prague},
addressline={Technicka 2},
city={Prague 2},
postcode={16627},
country={Czech Republic}}
\affiliation[2]{organization={School of Engineering, FEPS, University of Southampton},
city={Southampton},
postcode={SO17 1BJ},
country={United Kingdom}}
\fntext[fn1]{Corresponding author}
\fntext[fn2]{Corresponding author}

\begin{abstract}
Surface Enhanced Raman Spectroscopy (SERS) is a highly sensitive and selective technique that greatly enhances the signal of an analyte, compared with its signal from classical Raman Spectroscopy, due to its interaction with a substrates surface.
It has been shown that low concentration boron-doped graphene (B-graphene) enhances the Raman signal of simple organic molecules like pyridine.
Recent studies also suggest that B-graphene can remain thermodynamically stable when doped with significantly higher concentrations of boron than previously observed.
In this framework, we use quantum mechanical simulations to investigate the influence of dopant concentration and geometric distribution on the effectiveness of B-doped graphene as a SERS substrate, with glucose as analyte.
By combining analysis of interatomic force constants and of phonon eigenvectors composition, we conclude that higher doping concentrations provide a larger enhancement to glucose's Raman signal, while the molecule orientation relative to the surface plays a fundamental role in the Raman response.
We suggest that high concentration B-graphene presents itself as a potential substrate for SERS based detection of glucose, while the used phonon-based analysis can be promptly applied for the search of promising candidates as substrate materials for enhanced Raman response.
\end{abstract}

\end{frontmatter}

\section{Introduction}
\label{sec:intro}
Diabetes mellitus is one of the most prominent metabolic disorders of our time \cite{Zhou2024};
being characterized by chronic hyperglycaemia, it is usually diagnosed based on the fasting plasma glucose level.
The golden standard of testing for blood glucose have been electrochemical sensors since their inception \cite{D0CS00304B}.
Blood is commonly extracted using finger sticks or by drawing from a vein, both of which are invasive techniques that cause a certain degree of discomfort to the patient, while frequent blood glucose level tests are also necessary to manage the disease once it has been diagnosed. 
There is then a constant stimulus to create faster, easier, cheaper and ideally non-invasive methods for measurements of glucose levels. 
Researchers have proposed using microneedles for blood extraction or switching to analyzing sweat or the interstitial fluid of skin; 
however, small amount of collected blood or low glucose concentration in fluids other than blood leads to a decrease in accuracy when electrochemical sensors are used \cite{doi:10.1021/acs.analchem.8b03420}. 
One of the potential solutions is to develop sensors based on surface-enhanced Raman spectroscopy (SERS).
SERS is a highly sensitive and selective technique which greatly enhances the signal of an analyte compared to classical Raman spectroscopy, due to analyte interaction with a substrate \cite{CHEN2020115983}.
It is then a promising method to detect glucose from small blood amounts or bodily fluids with low glucose concentrations.
Research on the creation of SERS based glucose sensors is ongoing \cite{SUN2022339226}.
The most commonly used SERS substrates are noble metals, as they usually provide the largest Raman signal enhancement;
however, they have drawbacks such as poor glucose adsorption on bare surfaces \cite{doi:10.1021/ja028255v}, high monetary cost, adverse metal-adsorbate interactions, catalytic and photobleaching effects \cite{CHEN2020115983}.
Significant effort has been put into the research of alternative substrates, especially 2D nanomaterials such as graphene \cite{C5AN00546A}, transition metal dichalcogenide monolayers and Janus structures \cite{CHEN2020115983}.
In this study, we examine the ability of B-graphene to act as a SERS substrate for glucose detection using quantum mechanical simulations and phonon decomposition analysis.
We choose B-graphene as opposed to TMDs or Janus structures due to graphene's unique abilities such as suppressing fluorescence \cite{CHEN2020115983}. 
B-graphene has been shown to enhance the Raman signal of a simple analyte like pyridine by an order of 10$^3$ to 10$^4$ \cite{C2JM32050A,C2CP42297B};
such enhancement is significantly larger compared to graphene one by a factor ranging from 2 to 17 \cite{CHEN2020115983}.
However, these studies have only considered low doping concentrations and non-periodic boundary conditions.
The latter choice in the computational setup may introduce a severe approximation, which results in poor estimations of the substrate response.
In addition, when graphene flakes are used as a geometric model, the flakes are terminated with hydrogen atoms for the purpose of stability, which introduces fictitious interactions not present in the real system.
Lastly, to our best knowledge, no computational study on interactions between glucose and B-graphene exists.
For these reasons, the goal of the present work is to explore the effects of different concentrations of boron in B-graphene on the enhancement of the Raman signal of glucose.
We show that 12.5$\%$ concentration B-graphene provides a large enhancement to the Raman signal and, as such, it might be better suited as a SERS substrate for glucose sensing.
We also discuss how the relative orientation of the molecule with respect to the surface affects the Raman activity
Our phonon-based analysis is general and can be promptly applied for the prediction and characterisation of Raman signals in the presence of substrates, regardless the chemical composition of analyte and substrate.
\section{Computational details}
\label{sec:method}
\begin{figure}[ht!]
\centering
 \includegraphics[width=0.99\textwidth]{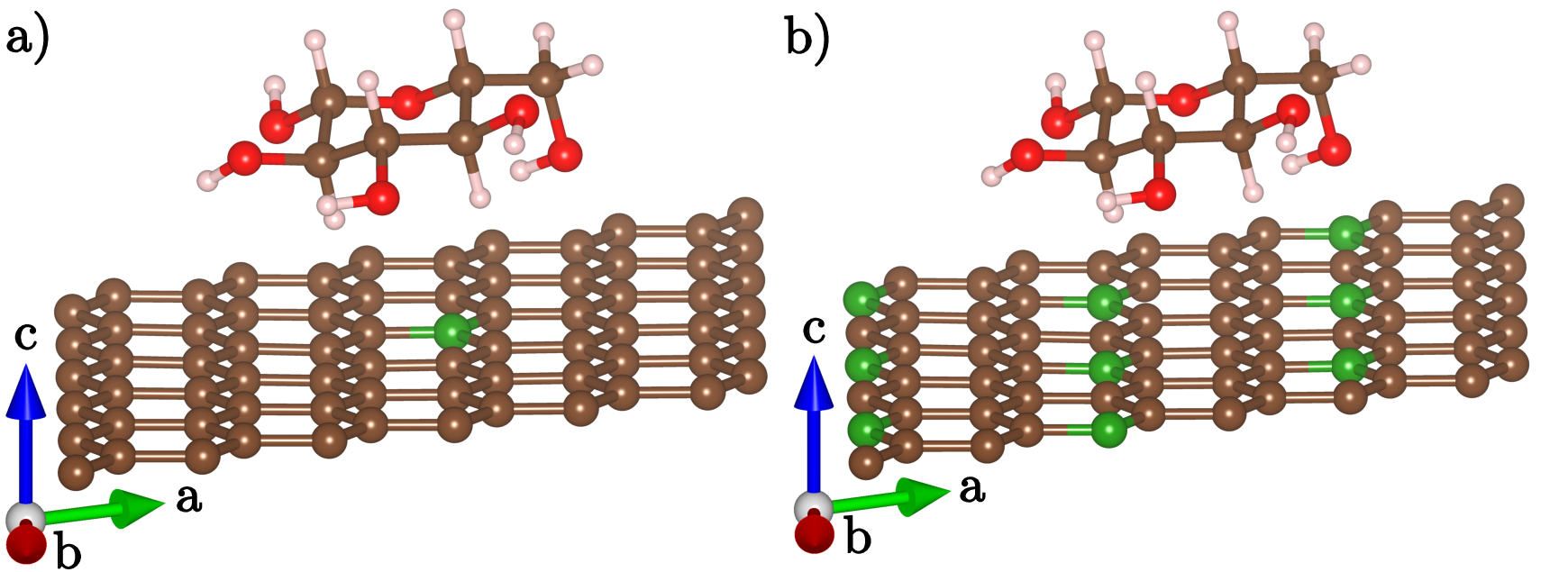}%
 \caption[Initial geometries of the studied systems]{Initial geometries of the studied systems (a) gLowBG, glucose on $1.39\%$ B-graphene, (b) gHighBG, glucose on 12.5$\%$ B-graphene.
Brown spheres represent carbon atoms, while red, white and green spheres represent oxygen hydrogen and boron atoms, respectively.}
 \label{fig:initgeo}
\end{figure}

We select $\beta$-glucose as analyte, as it is the most prominent anomer in aqueous solutions (e.g. blood) \cite{ARAUJOANDRADE2005143};
the reference atomic geometry is the conformer $^{4}C_1$ \cite{doi:10.1021/ja410264d}.
The graphene layer is placed in the $(\bm{a},\bm{b})$ plane of the unit cell, with the $\bm{c}$ lattice vector orthogonal to the plane and length set to 50 \r{A}, providing a vacuum slab large enough to avoid any interactions between replicas arising from the periodical treatment of the unit cell.
The substrate is modeled by a $6\times6\times1$ graphene supercell, on top of which we place the glucose molecule with its hydroxymethyl (i.e. CH$_2$OH) group pointing towards the substrate.
This is our reference system, which we name gG (glucose on Graphene).
We then modify the gG system by substituting carbon atoms of the graphene layer with 1 and 9 boron atoms, thus creating the gLowBG and gHighBG systems, respectively, the latter consistent with the reported stable geometry\cite{C6NR05309B} (\autoref{fig:initgeo}).
The graphene layer in gLowBG and gHighBG systems has a boron concentration of $1.39\%$ and $12.5\%$, respectively.

We perform ab initio calculation within the density functional theory (DFT) framework, as implemented in the VASP software \cite{PhysRevB.54.11169}.
We use the Perdew-Burke-Ernzerhof energy functional \cite{PhysRevLett.77.3865} together with the vdw-DFT-D2 van der Waals correction \cite{https://doi.org/10.1002/jcc.20495}, as suggested by previous studies \cite{JOSA2013170,doi:10.1021/ct800308k}.
Based on our benchmarks, we set the energy cutoff of the plane-wave basis set to 520 eV and the convergence criterion for the electronic Self-Consistent Field procedure to $10^{-8}$ eV, while the reciprocal space sampling is performed with a maximum of $7\times7\times1$ Monkhorst-Pack mesh\cite{PhysRevB.13.5188}.
In all systems including a glucose molecule adsorbed on top of the substrate, the length of the lattice parameters $\bm{a}$ and $\bm{b}$ correspond to a $6\times6\times1$ supercell of the primitive graphene cell.
Geometric optimization of the pristine graphene model is performed by varying atomic positions together with lattice parameters $\bm{a}$ and $\bm{b}$, while keeping the $\bm{c}$ lattice vector fixed.
For all the other models only atomic positions are relaxed while the lattice parameters are held fixed;
we make this choice because we assume that the substrate is adhered on a rigid surface for practical applications.

All the model geometries are relaxed until forces components on the atoms are less than $10^{-3}$ eV/\r{A}.
The geometry of the optimised structures is reported in \autoref{app:geo}.
Second-order force constants are calculated by means of Density Functional Perturbation Theory and postprocessed with the aid of the \textsc{phonopy} \cite{Togo_2023} software.
Raman spectra are evaluated by computing high frequency dielectric tensors at atomic geometries displaced along the Raman active phonon modes as reported in Ref. \citenum{C7CP01680H} with the aid of the \textsc{phonopy-spectroscopy} processing tool.

\section{Results and discussion}
\label{sec:result}
\subsection{Geometry optimization}
\label{sec:geom}

\begin{figure}
\centering
 \includegraphics[width=0.99\textwidth]{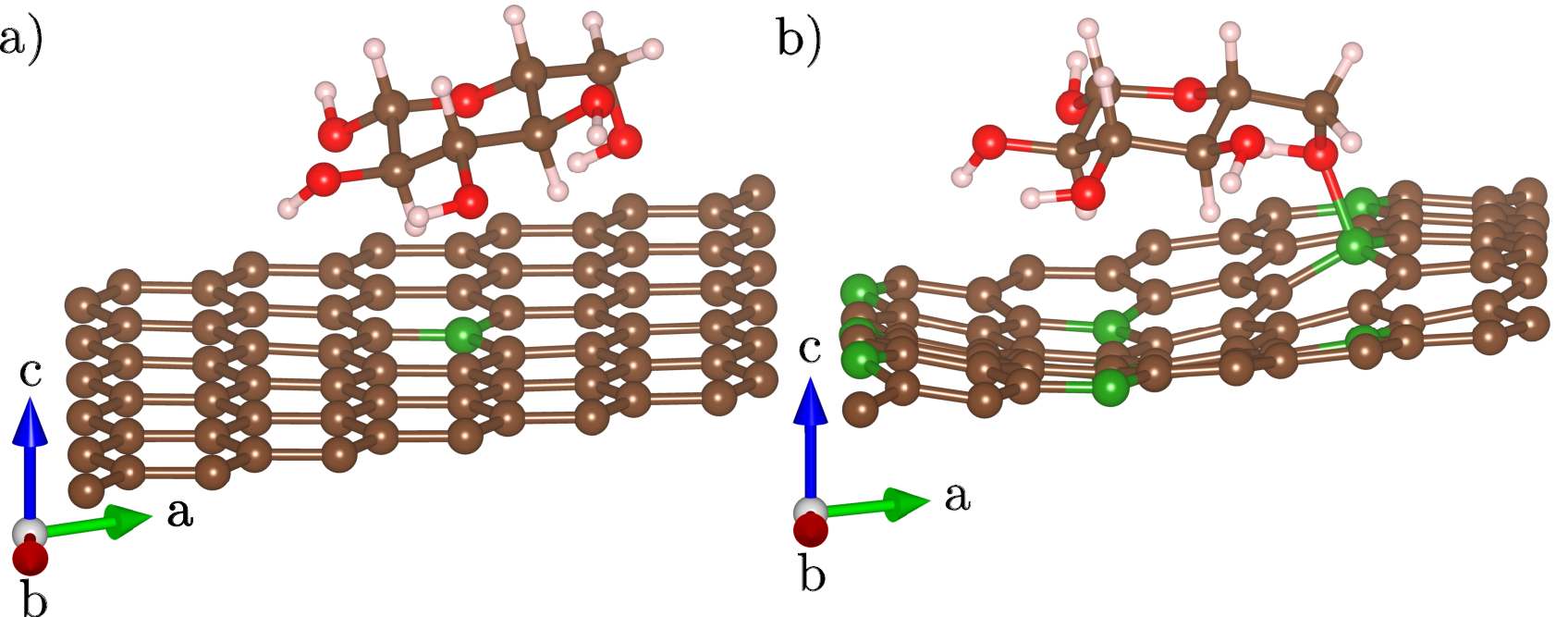}%
 \caption[Relaxed systems gLowBG and gHighBG]{Relaxed geometries of systems (a) gLowBG - glucose on $1.39\%$ B-graphene and (b) gHighBG - glucose on $12.5\%$ B-graphene.}
 \label{fig:gBGRelax}
\end{figure}
In the relaxed gG system (\autoref{fig:gBGRelax}), the binding distance of $3.140$ \AA{} between oxygen in the glucose hydroxymethyl group and its closest substrate atom agrees with previous studies of glucose adsorption on graphene \cite{DEVIP2023114134}.
A similar bond length of $3.116$ \AA{} is realized in the gLowBG case.
However, in the gHighBG system, glucose bonds with the substrate at a much shorter distance of $1.684$ \r{A}.
To quantify and compare interaction strength of glucose and substrate in the glucose/substrate systems we conduct force constant analysis on the glucose hydroxymethyl group oxygen and its closest substrate atom.
As the interatomic force constants of a couple of atoms constitute a $3\times3$ matrix, we compare absolute values of matrix traces divided by 3, in order to consider the average magnitude of a force constant on the main diagonal.
We consider the trace as relevant quantity because it remains invariant under spatial transformations of the system coordinates, unlike the full matrix, and thus contains the relevant information on the bond strength.
The trace average is found to be $\sim 35\times10^3$ times larger in gHighBG compared to the gLowBG case, indicating a much stronger glucose/substrate bond in gHighBG.
For the gLowBG and gG systems, the trace average is $\sim$6 times greater in the former compared to the latter, meaning that there is no significant difference in interaction strength between the two systems.
To check whether a glucose/substrate bond similar in strength to the gHighBG case can form on low concentration B-graphene, we consider the following system.
We position glucose on top of $1.39\%$ B-graphene so that the hydroxymethyl group oxygen is within the same distance of the substrate boron as in the optimized gHighBG system.
During geometric relaxation, glucose is repelled to a distance similar to the one found in the gLowBG case.
This result suggests that glucose cannot form a strong bond with B-graphene if doped at low concentration.
\subsection{Phonons and Raman spectra}
\label{sec:ram}
For all the considered systems containing a glucose molecule, the only symmetry operation is the identity;
as such, all phonon modes at $\Gamma$ are Raman active \cite{Kroumova01012003}.
The frequency range of interest for studying the glucose Raman signal is [600,1600] cm$^{-1}$, as reported in Ref. \citenum{ARAUJOANDRADE2005143}.
Accordingly, we only compute Raman intensities of modes below 1600 cm$^{-1}$.
Because we are only interested in relative peak positions and intensities, we do not study their temperature-dependent finite width.

First, we calculate the spectrum of pristine graphene as a reference.
It consists of a single peak called the G band with a wavelength of 1576 cm$^{-1}$, corresponding to the first order Raman scattering process, the signal of which is reported to be at about $\sim$1580 cm$^{-1}$ \cite{MALARD200951};
therefore, we assume that the simulation setup causes a red shift of about $\sim$4 cm$^{-1}$ between the calculated and the expected experimental value.
To allow comparison with possible reported experimental data, from now on we report all the calculated spectra shifted by the same amount.
To compare the enhancement of the glucose Raman signal among our systems, we first need to understand which peaks originate from glucose and which from the substrate, and which peak in one system corresponds to a given peak in another system (eigenvector comparison);
we then need to identify the atomic character of the relevant phonon modes and compare their eigenvectors across the systems, respectively.
The atomic character quantifies whether the atomic displacement pattern of a mode mainly involves motions of the substrate or the molecule;
we calculate it by means of the \textsc{phonchar} code\cite{PhysRevB.103.035406}.
To establish a one-to-one correspondence between modes originating from the same species, we compare the phonon eigenvectors at varying systems with the aid of the \textsc{eigmap} code\cite{10.1063/5.0224108}.
Both the \textsc{phonchar} and \textsc{eigmap} software codes are available free of charge\cite{github}.
Characterization of phonon eigenvectors is necessary to track how the phonon frequency changes at varying conditions;
this has been shown, for instance, for the case of the evolution of the Raman signal in transition metal dichalcogenide hetero-bilayers as a function of temperature\cite{10.1063/5.0224108}.
\begin{figure}[ht!]
\centering
 \includegraphics[width=.99\textwidth]{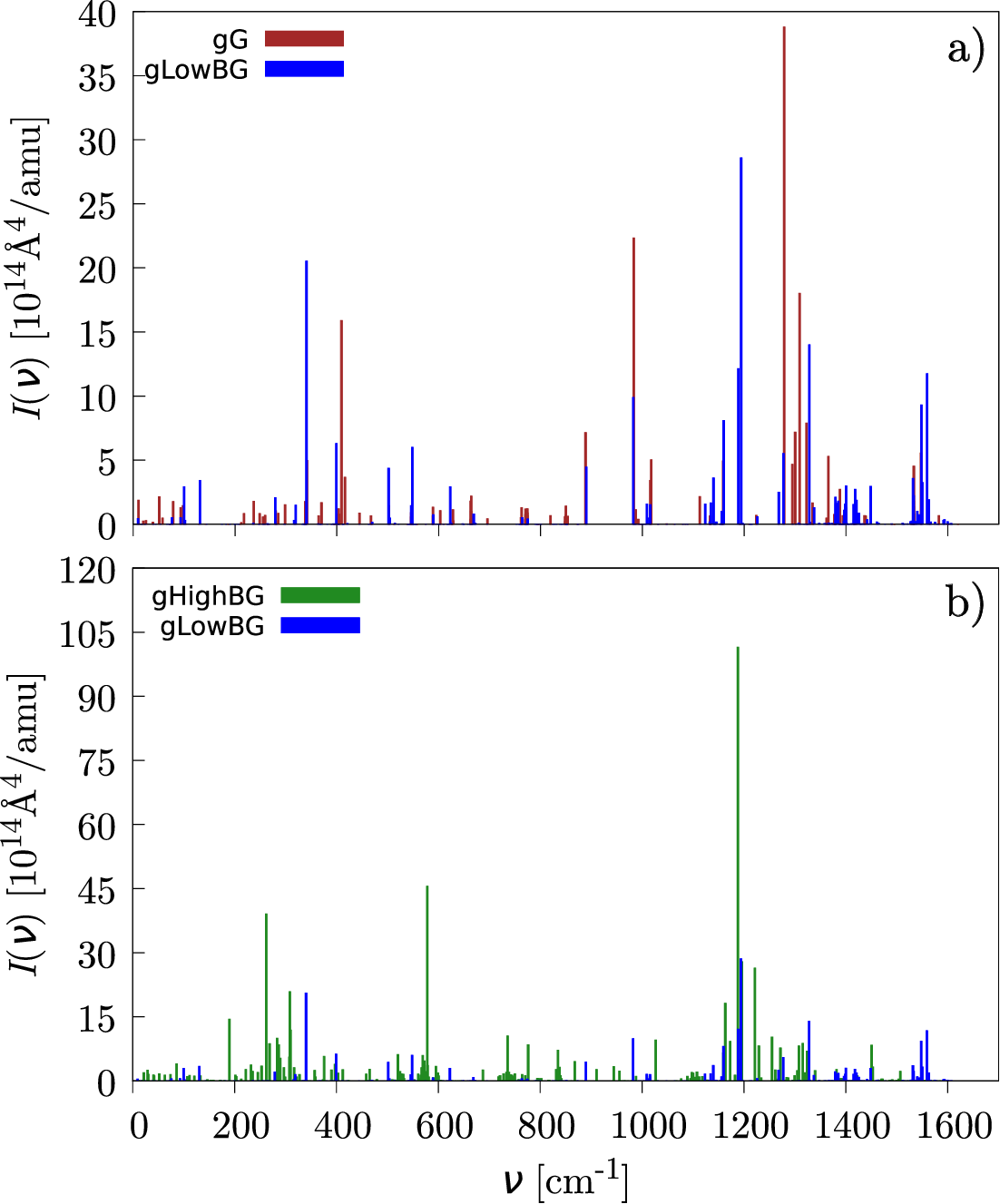}
 \caption[Raman spectra of systems gG, gLowBG and gHighBG]{Raman spectra of (a) gLowBG (glucose on $1.39\%$ B-graphene) and gG (glucose on pristine graphene) systems, and (b) gHighBG (glucose on $12.5\%$ B-graphene) and gLowBG systems.
 }
 \label{fig:gBGRam}
\end{figure}
We now compare the Raman spectra of systems gG with gLowBG and gLowBG with gHighBG (\autoref{fig:gBGRam}), being this the central result of the present work.
Thanks to the abovementioned phonon character analysis, we determine that in all the glucose/substrate systems, Raman signals in frequency range [870,1350] cm$^{-1}$ is mainly due to the Raman activity of glucose.
The peaks in region [1500,1650] cm$^{-1}$ represent the vibrations of graphene and B-graphene in the gG and gLowBG systems, respectively;
instead, in gHighBG, peaks corresponding to the vibrations of the substrate appear around 1451 cm$^{-1}$.
By comparing these spectra with the reported spectrum of $\beta$-glucose \cite{ARAUJOANDRADE2005143} and our calculation of it (see \autoref{app:gRaman}), we see that all substrates enhance the Raman signal of glucose by roughly $\sim$10$^{14}$, with significant wavelength shifts and modulation of the overall spectra.
The most intense glucose peak in the gG system appears at 1279 cm$^{-1}$, corresponding to the 1277 cm$^{-1}$ peak in gLowBG and 1287 cm$^{-1}$ peak in gHighBG;
their relative intensities to the corresponding gG peak are 1.4$\times 10^{-1}$ and 1.8$\times 10^{-2}$ for the gLowBG and gHighBG systems, respectively.
In the B-graphene systems the most intense glucose peaks have frequencies 1194 and 1188 cm$^{-1}$ in gLowBG and gHighBG and correspond to the 1186 cm$^{-1}$ gG peak. Their relative intensities are 1.1$\times 10^{4}$, 3.7$\times 10^{4}$ in gLowBG and gHighBG, respectively.
These specific examples illustrate that different peaks become prominent and each mode is shifted by a non-constant factor towards higher or lower frequencies;
overall, the $12.5\%$ B-graphene seems to provide the greatest enhancement.
A comprehensive list of the most intense glucose peaks and their relative magnitude is provided in Table 1 in \autoref{app:peaktable}.
A special attention should be paid to the peaks at 190 cm$^{-1}$, 578 cm$^{-1}$ and 1163 cm$^{-1}$ in the gHighBG spectrum.
These modes do not correspond to any modes in the other systems as they are peculiar of only the high concentration B-graphene system.

We can qualitatively compare these results to how B-graphene influences pyridine by considering what has been reported in literature \cite{C2JM32050A}.
They concluded that their low concentration B-graphene substrate ($\sim 2.4\%$) can increase the SERS signal of pyridine when compared with pristine graphene.
This does not seem to be true for glucose.
The pyridine molecule was more strongly bonded with the B-graphene than with the pristine graphene, as indicated by a significantly higher adsorption energy.
Hovewer, in our case there is not much difference in the interaction strength between glucose on pristine graphene or low concentration B-graphene, which could explain why these substrates provide a similar enhancement. 
In the case of the $12.5\%$ B-graphene there is a significant increase in the interaction strength and, as previously mentioned, it provides the largest signal enhancement. 
Consequently it may be better suited as a substrate for SERS based detection of glucose.
\begin{figure}[ht!]
\centering
 \includegraphics[width=0.99\textwidth]{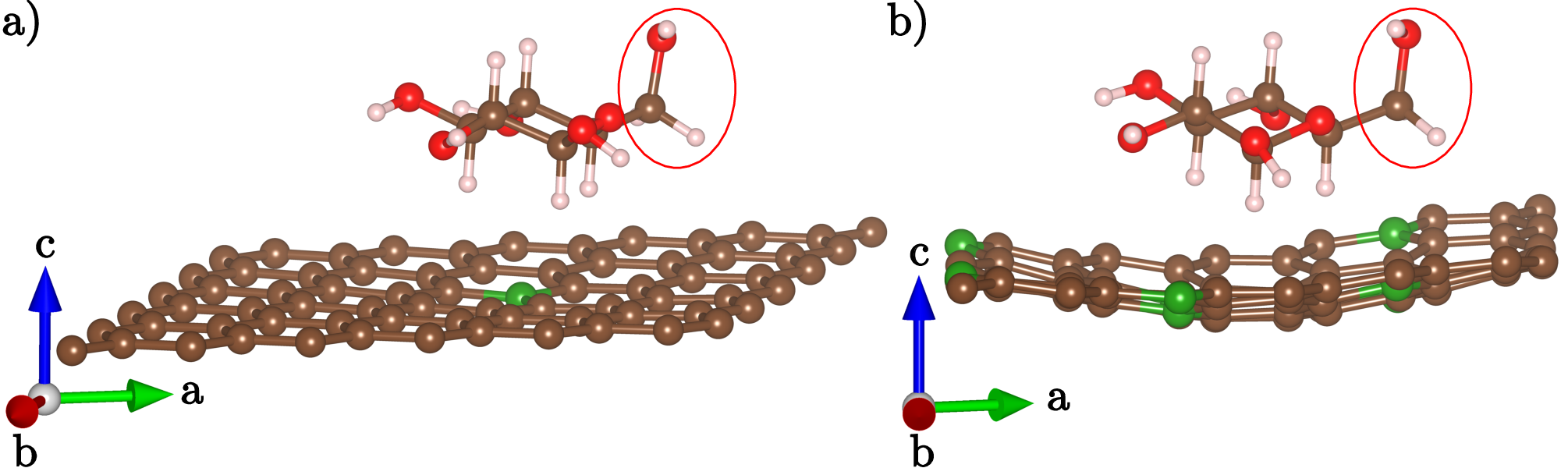}%
 \caption[Relaxed systems gLowBGF and gHighBGF]{Relaxed geometries of (a) gLowBGF and (b) gHighBGF systems, where the glucose molecule is flipped; i.e. its hydroxymethyl group, highlighted by a red outline, is pointing away from the substrate.}
 \label{fig:gBGF}
\end{figure}
To investigate the influence of the hydroxymethyl group's proximity to the substrate on glucose's signal enhancement, we decide to study 2 more systems (\autoref{fig:gBGF}), which we name as gLowBGF and gHighBGF.
They are built by considering the geometry of gLowBG and gHighBG systems and flipping the glucose molecule with the hydroxymethyl group pointing away from the substrate.
\begin{figure}[ht!]
\centering
 \includegraphics[width=.99\textwidth]{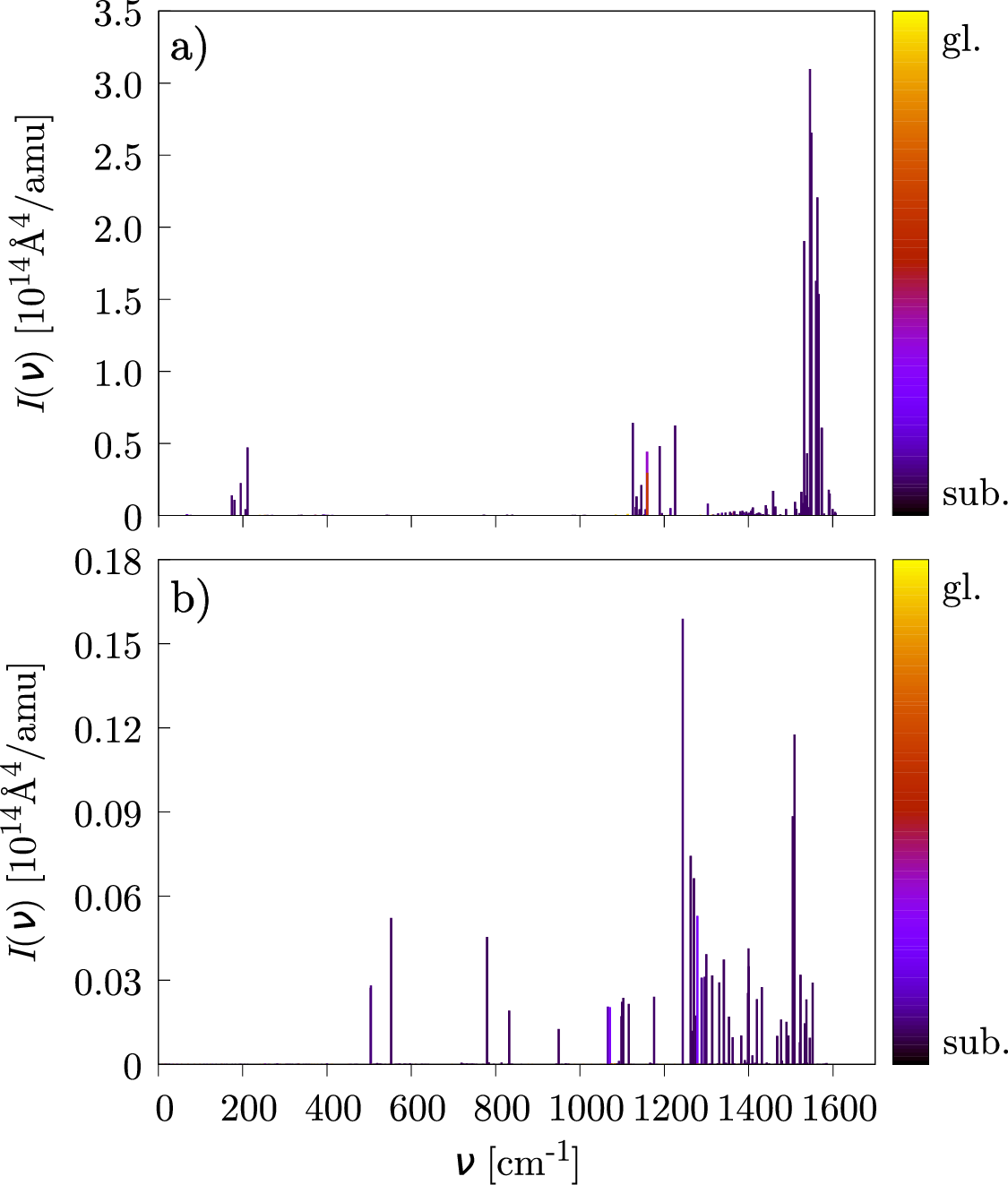}
 \caption[Raman spectra of systems gLowBGF and gHighBGF]{Raman spectra of (a) gLowBGF and (b) gHighBGF systems. Peak colour gradient indicates the atomic character of the phonon displacements: pure yellow and dark purple colours correspond to contributions from only the glucose molecule or the substrate, respectively.
}
 \label{fig:gBGFRam}
\end{figure}
Spectra of the gHighBGF and gLowBGF systems (\autoref{fig:gBGFRam}) underline the importance of the proximity of the glucose hydroxymethyl group to the surface in determining the signal enhancement.
In both cases, peaks corresponding to substrate vibrations become the most prominent and the Raman signal is overall weaker than in the gHighBG and gLowBG systems.
Glucose modes are, of course, still present, but weaker by several orders of magnitude than in the non-flipped cases.

\clearpage
\section{Conclusion}
\label{sec:conclusion}
We have studied the role of boron concentration in B-graphene in enhancing glucose's Raman signal using quantum mechanical simulations and phonon eigenvector analysis.
Our results suggest that 12.5$\%$ B-graphene provides an optimal enhancement to the glucose Raman signal compared to pristine graphene and to B-graphene systems with lower concentrations.
The large enhancement relates to the substrate's ability to from a stronger bond with glucose compared to the pristine and low concentration cases.
We then suggest that a dopant concentration equal to 12.5$\%$ as in the gHighBG system might be ideal when using boron-doped graphene as a SERS substrate for glucose detection.
The performed phonon eigenvector analysis turns out to be fundamental to correctly track the evolution of the Raman signals as a function of dopant concentration.
Such analysis can be promptly exploited in other studies on phonons evolution regardless the chemical composition and topology of the system of interest.

\section*{CRediT authorship contribution statement}
\label{sec:credit}
\textbf{Jan Komeda:} Conceptualization, Investigation, Formal analysis, Visualization, Writing - original draft, Writing - review \& editing.
\textbf{Antonio Cammarata:} Funding acquisition, Supervision, Writing - review \& editing, Methodology.
\textbf{Tom\'a\v{s} Polcar:} Funding acquisition, Supervision.

\section*{Declaration of competing interest}
\label{sec:declare}
The authors declare that they have no known competing financial interests or personal relationships that could have appeared to influence the work reported in this paper.

\section*{Data availability}
\label{sec:data}
Data will be made available on request.

\section*{Acknowledgments}
\label{sec:ack}
This work was co-funded by the European Union under the project ``Robotics and advanced industrial production'' (reg. no. CZ.02.01.01/00/22\_008/0004590).
This work was supported by the Ministry of Education, Youth and Sports of the Czech Republic through e-INFRA CZ (ID:90254).

\appendix

\section{Relaxed structures of the glucose/substrate systems}
\label{app:geo}
In this section, we provide the relaxed structures of the glucose/substrate systems in POSCAR format.
 \subsection{gG}
    {\small 
\begin{verbatim}
Relaxed geometry of glucose on a 6x6x1 supercell of pristine graphene                
   1.00000000000000     
    14.8064165343998191    0.0000000000000000    0.0000000000000000
    -7.4032082671999104   12.8227328578041888    0.0000000000000000
     0.0000000000000000    0.0000000000000000   50.0000000000000000
   C    H    O 
    78    12     6
Direct
  0.1131065875530136  0.0412458500623911  0.4989353602713414
  0.2806835883665624  0.0423852408426491  0.4989856658759601
  0.4476182376628596  0.0414896661247749  0.4990049225212315
  0.6131191946603383  0.0411870302574169  0.4989232959698648
  0.7806096260139830  0.0422942724860210  0.4988975847441822
  0.9475554946511517  0.0414830019550553  0.4989117835171469
  0.1140161291057721  0.2090505935704816  0.4986421643307774
  0.2808683719651340  0.2082032867357052  0.4984050820049853
  0.4465225326611377  0.2079245408458312  0.4984908624718579
  0.6139935494040821  0.2090253097343887  0.4987833169922462
  0.7809028751774644  0.2081457989888426  0.4989820084486276
  0.9464493758026399  0.2078954841544261  0.4989399087143947
  0.1143059722355763  0.3748532909216050  0.4985921097040076
  0.2797815229477265  0.3745841242489799  0.4977553052725116
  0.4473497853919522  0.3757634347241579  0.4973017954762481
  0.6142555043304440  0.3748790483965926  0.4979669087905635
  0.7797866846260832  0.3746083048076077  0.4987038063300024
  0.9473344956530688  0.3757005600085549  0.4989178197489093
  0.1131473389753655  0.5412564998732826  0.4988234506757746
  0.2806851374994294  0.5423353445948430  0.4981713908707509
  0.4475719819337514  0.5414928251908809  0.4972428513780340
  0.6132174500359362  0.5413010093669071  0.4971760962291675
  0.7806506959900170  0.5424492889377084  0.4981105051511173
  0.9475710936453503  0.5415333272156777  0.4988189954149758
  0.1140132268988092  0.7090478153815847  0.4990130218953561
  0.2809701928807261  0.7081809004260158  0.4987776719554831
  0.4465466005407878  0.7079402380288786  0.4980386980657799
  0.6140020605467162  0.7090760881973003  0.4976742800170357
  0.7808626216532937  0.7081210466950042  0.4980851396771134
  0.9464301091803311  0.7079149387064970  0.4987544656897253
  0.1142538366486452  0.8748429539062716  0.4990745152841708
  0.2798447030707245  0.8745925833372007  0.4990945433876591
  0.4474028212582248  0.8757179552931809  0.4988052548165594
  0.6142392893424803  0.8748006630662762  0.4984107128567019
  0.7797340239457298  0.8745317501497251  0.4985460548621393
  0.9472829669916782  0.8756633867426766  0.4988446051066464
  0.0577706126954101  0.0966031305197374  0.4988713053500842
  0.2255533346756815  0.0974816507215307  0.4988331635825859
  0.3914036774472515  0.0977147459223141  0.4988904328548477
  0.5577942001516725  0.0965720975301373  0.4989576899834509
  0.7255345949802935  0.0974206990593060  0.4989534283726639
  0.8913271803300222  0.0976888795225041  0.4989385036116662
  0.0588965057903159  0.2641302404823277  0.4987430625102559
  0.2247105337667765  0.2644677764135753  0.4982091843345418
  0.3911223621047167  0.2632911473262488  0.4981127932129905
  0.5589466042897049  0.2641731458301541  0.4984033183662455
  0.7247050770576756  0.2644004188726642  0.4988498922576619
  0.8911051416625255  0.2632581541585364  0.4989683744604410
  0.0580620167465546  0.4310641443787164  0.4988121726267802
  0.2244959018299461  0.4299195904725346  0.4981275342035691
  0.3921418299843994  0.4307794359145483  0.4972239793002430
  0.5580935517172535  0.4311120213310544  0.4972600769396078
  0.7244534437340031  0.4300185193196410  0.4982623673796999
  0.8922333429975616  0.4308216274743469  0.4988388430742888
  0.0577873767997525  0.5966136543759987  0.4989242513247999
  0.2256061203751899  0.5974529503228987  0.4986191236957174
  0.3913803913993787  0.5977078096275188  0.4978073426558314
  0.5578055022705234  0.5966281225010084  0.4971971816634385
  0.7255815957500233  0.5974983065068840  0.4976893041604208
  0.8913190112871812  0.5977319900427168  0.4985888461451133
  0.0588715282925712  0.7641230124649202  0.4990068642507559
  0.2247278294973669  0.7643947535337987  0.4990285786850561
  0.3911755435752728  0.7633054862848015  0.4985795560841119
  0.5589876190208594  0.7641852668779991  0.4979355856792599
  0.7246805047592143  0.7643598607801949  0.4980838438562313
  0.8910573516908461  0.7632289867432467  0.4986023479399862
  0.0580098921140540  0.9310427191475217  0.4990009006320863
  0.2244681156136216  0.9299323360512224  0.4991186590454039
  0.3922780007270409  0.9307795103263233  0.4990206086301494
  0.5580632973329732  0.9310010692549310  0.4987330472116711
  0.7243719800609303  0.9298606278695580  0.4986293560023116
  0.8921686692280127  0.9307562625538660  0.4988086609560992
  0.6017735571457955  0.5344243334720372  0.5752742658992479
  0.4997331901093460  0.4565226471183387  0.5619607042486310
  0.4090768399380377  0.4580477629530612  0.5761647325575181
  0.5172602667014716  0.6380434002372182  0.5882033263071575
  0.6130605461553670  0.6425358227818379  0.5750493190316457
  0.5173670905203992  0.7407888360600272  0.5862007315038443
  0.6758188757860759  0.4703352957668723  0.5589306454985522
  0.5157506338847022  0.6176613302382413  0.6096068000998531
  0.4999970684893931  0.4810649056764439  0.5411855653835461
  0.6008762835668050  0.5104780499514406  0.5963354365892277
  0.4051018167773263  0.4347225796859428  0.5974503978304661
  0.4368991857500955  0.3109105930520265  0.5507432813971606
  0.2580039847193197  0.3820910994771436  0.5726134983727279
  0.4498188836881619  0.7333405790926774  0.5975267180356826
  0.4563926110507930  0.7123819859446783  0.5504451823276739
  0.6177014190579302  0.6682208741995179  0.5540703560569307
  0.5895503175821314  0.8026966312111723  0.5952213138723614
  0.7614165598531889  0.7098617272897116  0.5829612479618482
  0.4233067746208389  0.5611655730479054  0.5750521998575243
  0.4925613821285862  0.3566970375864990  0.5629125978807790
  0.3174999228490300  0.3905291168412556  0.5623094319061919
  0.6901110848269250  0.5415481220619277  0.5616215322520133
  0.7028770236044362  0.7157039119290113  0.5897773291400309
  0.5157073229497972  0.7709247322090821  0.5591720025642172
\end{verbatim}}
 \subsection{gLowBG}
    {\small
\begin{verbatim}
Relaxed geometry of glucose on low concentration B-graphene
   1.0
    14.8064165343998191    0.0000000000000000    0.0000000000000000
    -7.4032082671999095   12.8227328578041888    0.0000000000000000
     0.0000000000000000    0.0000000000000000   50.0000000000000000
C H O B
  77   12    6    1
Direct
  0.1079507984602704  0.0368404169231518  0.4979686642644892
  0.2749982019030591  0.0374483221600664  0.4978519942812109
  0.4421721752251146  0.0373997402508879  0.4977146998315235
  0.6080234513180585  0.0369228094964928  0.4975766661796196
  0.7750051178600309  0.0375976379637765  0.4977270132587553
  0.9420168265901615  0.0373953682433043  0.4979613930298872
  0.1076401550731017  0.2035450386338859  0.4973520844490729
  0.2749393368970635  0.2032289804764376  0.4966924832863158
  0.4414201202781537  0.2027931984566343  0.4967432645970141
  0.6084407254172352  0.2044291588763152  0.4972407417786087
  0.7753959298032050  0.2040188921462674  0.4977327729557381
  0.9414732153019192  0.2037829624013606  0.4978305171767198
  0.1092358302969453  0.3709351729975834  0.4969950562775019
  0.2736268777854396  0.3682744538719706  0.4952967832918060
  0.4408426643719290  0.3688075351151431  0.4944841868132249
  0.6085425912020374  0.3709752150312401  0.4957865267398246
  0.7746697280353585  0.3704115838912325  0.4972350175074905
  0.9418670438415910  0.3710873296627729  0.4977017723950883
  0.1072634606029846  0.5371492165480179  0.4974659076942081
  0.2739201309515770  0.5387106855208094  0.4958411072269010
  0.6103560720754945  0.5380720397482034  0.4944217847329874
  0.7760280154146171  0.5378201552490887  0.4964295582635546
  0.9420137581282850  0.5373496697376755  0.4976507489812156
  0.1083363871726676  0.7045415727714361  0.4979837652946431
  0.2750422745529090  0.7045353923557133  0.4971586542760131
  0.4405149296524321  0.7051001200224165  0.4955601392258998
  0.6105432377887190  0.7055464017423074  0.4951826707338321
  0.7761527721973731  0.7043940912527661  0.4963909173648675
  0.9416456603374144  0.7038766955751028  0.4976889204620287
  0.1087405720388919  0.8707214638642107  0.4981998802841168
  0.2746470806874719  0.8705659523346717  0.4979869799018196
  0.4417619977002349  0.8719089033435279  0.4972722092293387
  0.6085385261489619  0.8702449447450273  0.4967067116657847
  0.7755763020828544  0.8712393342815723  0.4971783960856053
  0.9419694866567391  0.8711308866536593  0.4978772694386732
  0.0526493088713953  0.0922016718197983  0.4978739669301954
  0.2199551917848387  0.0923223087409383  0.4976279765463991
  0.3864457432765386  0.0927833976020459  0.4975503567498085
  0.5529196868961164  0.0925636757541394  0.4975807280409089
  0.7201961393563590  0.0928558170990191  0.4977069263453608
  0.8865326439091245  0.0932025873113457  0.4979037263330312
  0.0533817867495446  0.2592798955338384  0.4974449501837434
  0.2186253940313271  0.2583631602351996  0.4964217248592026
  0.3859955091868621  0.2577135426625748  0.4960081134848771
  0.5535146706794659  0.2593153879302563  0.4965492960716684
  0.7198834089749244  0.2599249545608740  0.4974176632792713
  0.8863197576761951  0.2592796408235006  0.4977924435124834
  0.0530618398127850  0.4267141397287371  0.4974554653896235
  0.2193102658788115  0.4257346874334159  0.4959379583430059
  0.3842031464070270  0.4210503100808932  0.4940123676219998
  0.5535082982034421  0.4264015770968864  0.4945340117044134
  0.7199515092502278  0.4261262174683770  0.4964713546476585
  0.8869065589291906  0.4263149009891162  0.4975790987330115
  0.0526191574815850  0.5927358251176136  0.4978049846089327
  0.2189584486130828  0.5933046793902865  0.4968813983575150
  0.3839959346673986  0.5958517228841962  0.4949288171142889
  0.5581150526822110  0.5952002189774059  0.4940627954453704
  0.7217595922300341  0.5932179101949010  0.4956320875183269
  0.8869130967477536  0.5933865567433578  0.4973416184296034
  0.0535733110739822  0.7596780748238834  0.4980910719662594
  0.2195840812832092  0.7602880652553635  0.4978414374356935
  0.3859704074179029  0.7606341783518231  0.4966953242432098
  0.5538372254004105  0.7600383450786857  0.4956420782638513
  0.7214025235932829  0.7611449064562155  0.4962495832014264
  0.8868321883624108  0.7595153637923369  0.4974241405512231
  0.0532226376694917  0.9265337734717802  0.4981417112128947
  0.2195316298170244  0.9258954288872381  0.4981524336397388
  0.3868218025146041  0.9265780397042479  0.4977571485793717
  0.5531069933150449  0.9267079486222255  0.4972156195981276
  0.7198817409586349  0.9260801973625166  0.4972434269792246
  0.8872381536202866  0.9265122061285049  0.4977797663014581
  0.5934353151411730  0.5311924214033256  0.5723943502605339
  0.4939956669191393  0.4487941992012273  0.5589171060338040
  0.4000362473722580  0.4418637839419419  0.5735225421233001
  0.4970001069915388  0.6229083408410651  0.5856082625853186
  0.5960888487816824  0.6354052609374279  0.5723382459749904
  0.4881183422767925  0.7208254457959785  0.5835366036148536
  0.6749469544661005  0.4763017857084718  0.5555729872446972
  0.4976212653698646  0.6036356189719844  0.6070186786556240
  0.4916974845862993  0.4732302037137466  0.5382556438418470
  0.5941480276207571  0.5075706233043594  0.5934130448249825
  0.3993282359578072  0.4195315319597585  0.5947819654578159
  0.4418600637603918  0.3056277930487836  0.5468778175869730
  0.2493822823696062  0.3564543258615877  0.5706189151200318
  0.4179907363175003  0.7074833916198695  0.5946962815539333
  0.4311319734251248  0.6875549705645337  0.5473342912807446
  0.5986096499503502  0.6604901631732584  0.5513859636882100
  0.5574414698041834  0.7861303928791382  0.5927130186473796
  0.7445846025215456  0.7116422793195276  0.5803149449254510
  0.4055617095049837  0.5413330506094907  0.5726016751871018
  0.4950542120106696  0.3528072378283641  0.5594068904906667
  0.3097693208583641  0.3691860156319556  0.5600267324657261
  0.6849083205047302  0.5456740312717072  0.5588203547838505
  0.6834145931988377  0.7127187959560026  0.5870901038936539
  0.4845557952590764  0.7496369024615316  0.5564846197679933
  0.4419935553321508  0.5373896537661760  0.4937453483507556
\end{verbatim}}
 \subsection{gHighBG}
    {\small
\begin{verbatim}
Relaxed geometry of glucose on high concentration B-graphene
   1.0
    14.8064165343998191    0.0000000000000000    0.0000000000000000
    -7.4032082671999095   12.8227328578041888    0.0000000000000000
     0.0000000000000000    0.0000000000000000   50.0000000000000000
C H O B
  69   12    6    9
Direct
  0.2991280223704047  0.0671156708208167  0.4983474397760221
  0.6313339074163467  0.0661546678793724  0.5085353181406086
  0.9665737210029781  0.0682088277465155  0.5033496528286410
  0.1285970546310133  0.2318780579051445  0.4930273221020915
  0.2991766029340442  0.2317199363709890  0.4890519661671123
  0.4634342150033195  0.2302832749169951  0.4897897429312169
  0.6319717692337902  0.2303500374607240  0.4974238417346858
  0.7960892246181878  0.2312244789586540  0.5034833764625068
  0.9664621053018336  0.2324067925692209  0.5020600501520018
  0.1322458440775092  0.4019970249530507  0.4958093896208767
  0.4664445420556598  0.4027374589543565  0.4842878314451536
  0.7989757563740653  0.4012085004193962  0.4976823883648482
  0.1326679415646487  0.5657795064250239  0.5039898742572704
  0.2969912019121809  0.5662809698384669  0.4979919650136936
  0.4657578774690300  0.5677577128940315  0.4915730261597563
  0.6312742631859908  0.5678642571104255  0.4905519073049422
  0.8001042580781536  0.5664171410159426  0.4956270688428094
  0.9626254073408109  0.5653998460548310  0.5037236592726829
  0.2992205675432060  0.7345257693904644  0.5098025942301616
  0.6327635573509836  0.7353003600705900  0.5021802314968014
  0.9656656756134485  0.7354827016434772  0.5048200814052860
  0.1305947437126231  0.8993505425250204  0.5058207400800079
  0.2987683322358019  0.8996209133012941  0.5088266278618547
  0.4631349858049044  0.8975160919870396  0.5114829800205990
  0.6322341430808939  0.8966729168759143  0.5122313331583479
  0.7976376972953679  0.8992944722099531  0.5065459787778543
  0.9659420870961576  0.8994297187405950  0.5052441969297453
  0.0727598545176103  0.1235580204925483  0.4968589375869740
  0.2475654620367051  0.1240834103523027  0.4951283069288808
  0.4074399743032718  0.1219609612943021  0.4954250540843159
  0.5796489264881031  0.1220439440927689  0.5026868778259412
  0.7406038212447354  0.1232943015489202  0.5085840928632116
  0.9148262826815660  0.1244551032221794  0.5064608382743053
  0.0758969927893693  0.2897857160529487  0.4962305006642537
  0.2390309397638254  0.2837509100509416  0.4879338714981316
  0.4105320582056221  0.2891668716868645  0.4862344410828207
  0.5744989103346678  0.2816017785687208  0.4907440917455456
  0.7428923127027386  0.2885656593051681  0.4991194906104098
  0.9065068658892927  0.2830329424977076  0.5021389246339255
  0.0804084399708744  0.4571185811338169  0.5004672408551041
  0.2399941735299492  0.4579558608618051  0.4921204759850766
  0.4130670937445858  0.4591784412731746  0.4851459164244272
  0.5751474785569682  0.4585485759321400  0.4862220421539869
  0.7477442883991910  0.4570722719332853  0.4929549258044504
  0.9061266160868395  0.4564781347873250  0.5020090979546665
  0.0727713833994105  0.6161847958943114  0.5066645889664049
  0.2439675134440620  0.6230616240900693  0.5048505454990911
  0.4067523425374332  0.6160320569110598  0.4985109513994977
  0.5773004546314159  0.6251944894044259  0.4946049471875857
  0.7413131191831474  0.6183530816501274  0.4942124990704607
  0.9096103246978871  0.6234991294374892  0.5017313532803473
  0.0741707746920709  0.7899254860337190  0.5077501341819670
  0.2449210562844238  0.7899532603750318  0.5105831857051970
  0.4074055841744770  0.7881901065635987  0.5122656484304449
  0.5797960996923556  0.7871549143397714  0.5100227823108304
  0.7421851289162689  0.7915658717614321  0.5014055451873943
  0.9118018174494052  0.7901818332630400  0.5037711514677432
  0.0753848818384952  0.9532072276868091  0.5042574628527308
  0.2430516666924785  0.9557907443920561  0.5047889886271456
  0.4080111481671982  0.9524554467049235  0.5080920690411391
  0.5756438167896875  0.9535209008548649  0.5119309850733025
  0.7425432470331544  0.9521641539225530  0.5112138955260541
  0.9104305061773643  0.9558401928952482  0.5053925948821281
  0.4898537846566901  0.4045047046302075  0.5614137871674889
  0.3757184069003336  0.3494684890413603  0.5526845657871888
  0.3168468231843694  0.3904481736093746  0.5693022459306963
  0.4778305465437543  0.5624458983271857  0.5725184585910184
  0.5366551643411357  0.5208735942581261  0.5562302253466120
  0.5106599227907406  0.6763738677991872  0.5656083808091570
  0.5062184136248472  0.2943953132718189  0.5447639487220723
  0.4936649956082164  0.5582344101282418  0.5939585909731526
  0.3707127895056843  0.3676008673788271  0.5315488712680272
  0.4938387149934254  0.3912047296425153  0.5831329016362138
  0.3163987309528014  0.3707907726551191  0.5906919612427625
  0.2820443605962988  0.2036445872647082  0.5424339418884535
  0.1720907490485393  0.3563889244365526  0.5720671067933426
  0.5060288834303448  0.7184510690962369  0.5831482756144206
  0.3761453629095330  0.6047260469014689  0.5467763758926336
  0.5270760123273254  0.5309673890087369  0.5347859583361025
  0.5893877751336570  0.7169888856233582  0.5570857365032804
  0.6821576301055735  0.5587267178670364  0.5523169561624085
  0.3675031993120077  0.5039402151473569  0.5672283887508625
  0.3336617606403559  0.2401424815882087  0.5564194736125926
  0.2169472225900470  0.3498985359366448  0.5589146186521990
  0.5492439380303639  0.3699367471984832  0.5467677362916370
  0.6433773202126749  0.5806843640217809  0.5637959571231059
  0.4375234653677831  0.6769250570642962  0.5459926027450291
  0.1324036129392404  0.0677342833389160  0.4984302560177548
  0.4648485448615650  0.0657008801738548  0.5016490099601469
  0.8005452480584117  0.0680621417333422  0.5113564227083197
  0.2965964683270285  0.4004697955827811  0.4858176998238266
  0.6330155256683102  0.3990511400494213  0.4882098827425976
  0.9641705886547538  0.3989189625798849  0.5016338645641445
  0.1299519383544588  0.7310841687677827  0.5095915223973967
  0.4617772517114602  0.7220035944731026  0.5143473727785017
  0.7982490587426084  0.7329496173277553  0.4987190825206355
\end{verbatim}}
 \subsection{gLowBGF}
    {\small
\begin{verbatim}
Relaxed geometry of flipped glucose on low concentration B-graphene
   1.0
    14.8064165343998191    0.0000000000000000    0.0000000000000000
    -7.4032082671999095   12.8227328578041888    0.0000000000000000
     0.0000000000000000    0.0000000000000000   50.0000000000000000
C H O B
  77   12    6    1
Direct
  0.1037941254190728  0.0285677032414646  0.4992004905249703
  0.2701116642580700  0.0288632026988554  0.4991087059907597
  0.4376401886216860  0.0294666737190814  0.4988094809725510
  0.6038786017277320  0.0287296212298722  0.4983948115431560
  0.7701602813435819  0.0290719773328690  0.4984319589388044
  0.9375574457654535  0.0295204617501474  0.4989321170768569
  0.1028174377004106  0.1949931878412799  0.4985282936850597
  0.2705207391840517  0.1952883135301190  0.4979611110687110
  0.4371607850501223  0.1945826432227995  0.4980507198772585
  0.6035333846744209  0.1958836852531885  0.4985310940241748
  0.7709260775317885  0.1961494000458950  0.4988882498586883
  0.9373595694190529  0.1955692989072369  0.4989084273215920
  0.1048069461424662  0.3631043102607506  0.4979928821753599
  0.2693958348673013  0.3599693714752910  0.4962082642709747
  0.4359140624461312  0.3601615843803135  0.4956155289430327
  0.6040606558161938  0.3630706986697256  0.4971613038588321
  0.7704692061224215  0.3621667844632687  0.4986691801108145
  0.9370205789333974  0.3625144902064988  0.4990021354286222
  0.1031401073745676  0.5288441659090887  0.4985561595860789
  0.2691418673494007  0.5301611787261636  0.4960894145698247
  0.6060388723158258  0.5297085909398760  0.4948938129083021
  0.7710667002096560  0.5291031130174981  0.4977473218208217
  0.9375423703425388  0.5293737984791167  0.4990830586341786
  0.1034793815718170  0.6959042546927887  0.4992368162985819
  0.2705259265752011  0.6965147571079046  0.4978165766801679
  0.4362120998969843  0.6967831090037514  0.4952806957938706
  0.6057618618850595  0.6970616110012978  0.4946876407613067
  0.7716809382122624  0.6964442373682503  0.4969477517955034
  0.9374403463879897  0.6955413285931237  0.4989607025840205
  0.1042516223129535  0.8627636957501644  0.4994748077415085
  0.2704434665364470  0.8622645939841065  0.4990742111391415
  0.4368347805440116  0.8633100143721699  0.4978222094134467
  0.6040151942292932  0.8623248216887388  0.4968361512219706
  0.7714530505894659  0.8630375619847899  0.4975241819751079
  0.9371372503857947  0.8625739688623396  0.4989028223049210
  0.0481181428201224  0.0846692207110696  0.4990229845934709
  0.2151250567045699  0.0841048652133366  0.4989168517457350
  0.3822671921022563  0.0849656916844591  0.4987809947557432
  0.5483730702478149  0.0850442995511552  0.4986569521763176
  0.7153886873280136  0.0847195318814542  0.4985970835637505
  0.8824034190226718  0.0854102940398770  0.4988251144691846
  0.0486229264175795  0.2511586633597224  0.4985499825810484
  0.2144839111735433  0.2504929409939416  0.4976342104780691
  0.3814043234654274  0.2500853434979663  0.4972681272105816
  0.5486364282069508  0.2511589224771862  0.4979416858968446
  0.7156979804930136  0.2521165793347664  0.4987318746223124
  0.8818174602446432  0.2517842692101060  0.4989765965810611
  0.0489269592082984  0.4188998095065640  0.4985987084524588
  0.2148005456259028  0.4182839149971471  0.4966329900704614
  0.3794202014151576  0.4128100682855374  0.4947296470077990
  0.5492657222835321  0.4184993631639463  0.4956303966790943
  0.7153651049437684  0.4185225856834016  0.4978809882982114
  0.8820699341273673  0.4180930998691188  0.4989969765164419
  0.0481406637037802  0.5851477901049834  0.4991094893120422
  0.2141955762893254  0.5850375838364059  0.4975355502701771
  0.3796775088536173  0.5879561692800896  0.4944852655995069
  0.5535910069526528  0.5875817256716210  0.4936792526575604
  0.7169084428668537  0.5849173731557770  0.4964221003991638
  0.8827048189855415  0.5854751870002930  0.4987361113756059
  0.0487592665970638  0.7514279492334829  0.4993946376449329
  0.2154708459209852  0.7523995338087061  0.4988715562651199
  0.3814521895187417  0.7529854460390485  0.4970055215547521
  0.5489193494448360  0.7518597446789692  0.4952984894340731
  0.7172121451108335  0.7533330391452755  0.4962922847013091
  0.8823163936361902  0.7519919901227573  0.4984164958606475
  0.0491041279782148  0.9186995442596976  0.4993710802070301
  0.2149953321556640  0.9183251115492934  0.4993882819039761
  0.3819435572049195  0.9183513774481112  0.4986955868242334
  0.5489132607292059  0.9189081420632845  0.4976800408104441
  0.7153797218437286  0.9186080971730201  0.4975764346717249
  0.8824460820967157  0.9183425328967999  0.4985845084481262
  0.4838165051836342  0.4697435553159118  0.5644183720297895
  0.5963061600277891  0.5277390691520717  0.5741271519202095
  0.6494977198718866  0.6388385966285228  0.5630455053163219
  0.4896193198915321  0.6402684547489460  0.5610132598280116
  0.4276164393013860  0.5287696627967503  0.5719332371697822
  0.4439743384161836  0.7087231949002811  0.5685482530417785
  0.4646655977070990  0.3314729046653432  0.5714543634167003
  0.4915127177498431  0.6358254493210457  0.5390172776074394
  0.5984680930160483  0.5321461906281332  0.5962131822257146
  0.4845492877687291  0.4649389387461426  0.5423446095275322
  0.6495920371934703  0.6364249792455386  0.5409490708883199
  0.7174853211472001  0.5093162010865306  0.5705006834864322
  0.7911140028137311  0.7563556877142136  0.5629640316922863
  0.4884845005456745  0.7840639794623260  0.5579314290800548
  0.5147722174076418  0.7490929208352035  0.6029144392780116
  0.4225161276282416  0.5317252966487230  0.5939450318005184
  0.3623920219036352  0.6703142092178020  0.5618652853550300
  0.2926914842377118  0.4068008148737761  0.5644777476253846
  0.5946523498129368  0.6907370494385628  0.5714829122634586
  0.6451867108529934  0.4713096566623212  0.5645481857226946
  0.7513815722813157  0.6925233397587595  0.5732512192895393
  0.4259133327434265  0.3669911447072142  0.5755394421148605
  0.3258834292567925  0.4807970259212364  0.5601592380459892
  0.4436553305163592  0.7238839034769216  0.5967427552587220
  0.4373536767359609  0.5292826396392200  0.4934865910360511
\end{verbatim}}
 \subsection{gHighBGF}
    {\small
\begin{verbatim}
Relaxed geometry of flipped glucose on high concentration B-graphene
   1.0
    14.8064165343998191    0.0000000000000000    0.0000000000000000
    -7.4032082671999095   12.8227328578041888    0.0000000000000000
     0.0000000000000000    0.0000000000000000   50.0000000000000000
C H O B
  69   12    6    9
Direct
  0.3011164469066131  0.0514451944445108  0.5050238296835524
  0.6338172550850146  0.0502499232401186  0.5046651339715296
  0.9678264648159131  0.0507563790286915  0.5100953417417904
  0.1318913795813379  0.2149360973997035  0.5005367066388564
  0.3010319518353440  0.2155162182795624  0.4971425961976161
  0.4647375591532204  0.2146183590401173  0.4948289182922710
  0.6346401851330267  0.2157557952761253  0.4996242591882763
  0.7985909854165170  0.2155091672436762  0.5049736067686676
  0.9688485648323587  0.2160913983425818  0.5067662655249328
  0.1350363236045789  0.3850374787257955  0.4964094272123108
  0.4678143012853718  0.3842620485673422  0.4862760056803089
  0.8016257876460501  0.3853722651459109  0.5001000812903272
  0.1350690828794517  0.5499809522375038  0.4983816462060963
  0.2984207638904479  0.5501143588982690  0.4890283194129329
  0.4683486588043633  0.5505555827937320  0.4836045220902745
  0.6343614961329990  0.5507701377943022  0.4887011711257747
  0.8021023814675381  0.5501517164381189  0.4975105396484435
  0.9661637136284735  0.5497112802399120  0.5033829847665918
  0.3019541371190245  0.7196739816198516  0.4986029863679916
  0.6364355438498821  0.7208554926473011  0.4931975645170073
  0.9674962552997645  0.7183215212702038  0.5085612139719176
  0.1327711353670195  0.8825280446994225  0.5118038930375414
  0.3014570614050451  0.8820123945381332  0.5069433680065687
  0.4659274494473511  0.8824412572968452  0.5008776673377425
  0.6341724257835396  0.8820310910869100  0.5012825317428149
  0.7979218121884651  0.8810940262654779  0.5058443876001628
  0.9666522860769408  0.8816151832358349  0.5118645177519890
  0.0764177491610411  0.1076872609170822  0.5067348885300959
  0.2499322307597994  0.1090483397036371  0.5046385652590014
  0.4087181347760329  0.1073023502067220  0.5005486134429569
  0.5824156881972050  0.1073091469266049  0.5022987341708675
  0.7420075468956868  0.1068549425013754  0.5071860619123243
  0.9160644086498430  0.1076443494298297  0.5108132116437449
  0.0793244696991507  0.2734820473154544  0.5009914320347578
  0.2412626389516387  0.2658159376485421  0.4946806580500513
  0.4115226106807971  0.2725751687677190  0.4918439679038392
  0.5759809667920655  0.2663943305616596  0.4943363582399581
  0.7459035589412101  0.2735477425601667  0.5012443931400422
  0.9094884219639490  0.2669694483940129  0.5052483151214083
  0.0826141094914924  0.4406572084222147  0.4987010367625050
  0.2415468142079089  0.4400637297262161  0.4903603724877121
  0.4146995550232587  0.4396919648286612  0.4833282718262515
  0.5771272723810909  0.4405084031847747  0.4872663251213435
  0.7499354495401672  0.4406830058557253  0.4951728214380955
  0.9092364852852285  0.4406305957164167  0.5031660786798582
  0.0763842303396339  0.6006249662138096  0.5028491216744313
  0.2460121375773307  0.6081644192756922  0.4949560575901971
  0.4086976482623080  0.6020838505820704  0.4852717957309151
  0.5806928462216199  0.6093977684454860  0.4874766418048306
  0.7435013011301574  0.6016384946450143  0.4947267601815953
  0.9121555283856160  0.6075760403087251  0.5035465690021773
  0.0767362818936400  0.7729111247963042  0.5108886101967554
  0.2487773517138093  0.7729747310051915  0.5053901997300555
  0.4104279126070933  0.7747647921442951  0.4959730854284364
  0.5823866692878030  0.7759082089469588  0.4938523364119786
  0.7433877538010252  0.7739480306057905  0.4997080525521395
  0.9126894522486626  0.7725400173735936  0.5087267888055889
  0.0770744424840387  0.9365032930426577  0.5130564712403669
  0.2454002299322611  0.9388705126614909  0.5088863374438354
  0.4104074100755467  0.9357698473464564  0.5046109637219993
  0.5778926915773595  0.9380794816809077  0.5024214761309226
  0.7418879999087217  0.9344054440644545  0.5069159375318020
  0.9103602761822391  0.9373486615418078  0.5101998337822321
  0.3404890525404053  0.4160596433688990  0.5588762072191210
  0.4385546780828434  0.4108558675830883  0.5650722027777966
  0.5299620794267221  0.5031973764144916  0.5510533123302662
  0.4529327447294022  0.6102087685893077  0.5548755054651524
  0.3535253764776323  0.5209640101074630  0.5673529536482010
  0.4782320376690446  0.7175832565585561  0.5647243382051265
  0.2458777505607505  0.2698277373748824  0.5677111380249024
  0.4420508352809076  0.6072708636497954  0.5329517559294271
  0.4537159515379698  0.4175603415570441  0.5868397132787908
  0.3286785155266498  0.4086021131092795  0.5368561706437617
  0.5130665940811964  0.4988338359549230  0.5294524720915377
  0.4870915153311359  0.3127399480659387  0.5575781324248154
  0.6715518259223476  0.5396488197612608  0.5419169230316777
  0.5455331950958072  0.7767877727945510  0.5531418689685715
  0.5537989428472486  0.7148034282130281  0.5962354707070646
  0.3602664615944011  0.5275948700979457  0.5893767348993966
  0.4106900859089005  0.7276595963025259  0.5609589656997347
  0.2046404550109090  0.4713426870694975  0.5648786592350382
  0.5417993872338429  0.5990076842822993  0.5612684464525318
  0.4220712270565721  0.3132168025623654  0.5550103734227073
  0.6209814146232426  0.4993981720021781  0.5558554380844538
  0.2498248139478692  0.3360087394904160  0.5719287155052929
  0.2677910810831824  0.5317399768111633  0.5581984821551367
  0.4982641066305914  0.7304709221767905  0.5927985389897249
  0.1355933174510624  0.0522176679779745  0.5095398132514081
  0.4673168561605966  0.0507940028421647  0.5029018474861786
  0.7997665075235523  0.0496357123619776  0.5108161885827496
  0.2982642643481306  0.3807062393160728  0.4873382280307094
  0.6348581685285877  0.3825796148815590  0.4907635077675002
  0.9676195642262555  0.3829991514324813  0.5029201630022396
  0.1337452614342742  0.7145631464491154  0.5073301073273219
  0.4671134965682891  0.7178753084231743  0.4895226913874239
  0.7997418233727825  0.7151037955037878  0.5010255230317974
\end{verbatim}}

\section{Calculated $\beta$-glucose Raman spectrum}
\label{app:gRaman}
\begin{figure}[ht!]
 \centering
 \includegraphics[width=.99\textwidth]{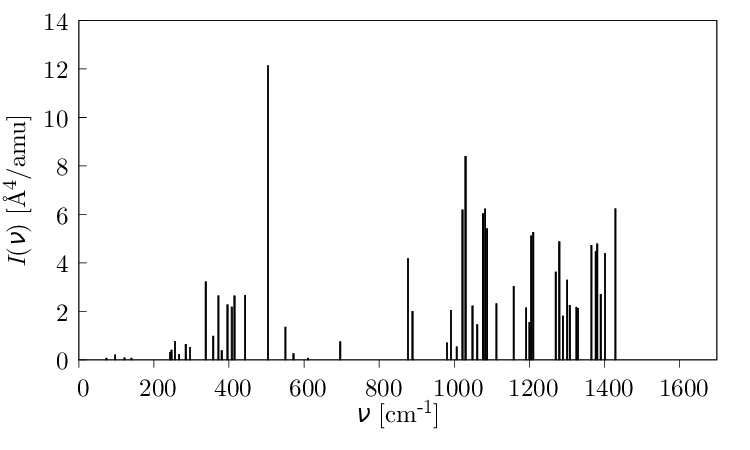}
\caption[Raman spectrum of $\beta$-glucose]{Raman spectrum of $\beta$-glucose.}\label{fig:gRam}
\end{figure}
The calculated Raman spectrum of $\beta$-glucose is shown in \autoref{fig:gRam}.
The spectrum is generally split into three main regions.
The anomeric region from 600-950 cm$^{-1}$ differentiates the $\alpha$ anomer from $\beta$-glucose with the latter having very few peaks in this area in the present study.
The region from $[950,1200]$ cm$^{-1}$ is called the \textit{fingerprint region}, while modes located in the range $[1200,1500]$ cm$^{-1}$ represent CH$_2$ and C$-$O$-$H vibrations.

\section{Glucose peak table}
\label{app:peaktable}
In \autoref{tab:peaks}, we summarize the prominent glucose peaks of the gG, gLowBG and gHighBG systems and their corresponding intensity ratio for a quantitive comparison.
\begin{table}[!ht]
    \centering
    \begin{tabular}{|c|c|c|c|c|}
    \hline
        $\nu_{gG}$ & $\nu_{gHighBG}$ & $I_{gHighBG}/I_{gG}$ & $\nu_{gLowBG}$ & $I_{gLowBG}/I_{gG}$ \\ \hline
        266 & 262 & $2.1\times10^{6}$ & 264 & $1.9\times10^{3}$ \\ \hline
        341 & 308 & $4.2\times10^{0}$ & 340 & $4.1\times10^{0}$ \\ \hline
        409 & 412 & $1.7\times10^{-1}$ & 421 & $3.5\times10^{-3}$ \\ \hline
        502 & 502 & $5.9\times10^{-2}$ & 502 & $3.0\times10^{0}$ \\ \hline
        549 & 536 & $4.5\times10^{0}$ & 548 & $3.7\times10^{3}$ \\ \hline
        888 & 868 & $6.3\times10^{-1}$ & 890 & $6.2\times10^{-1}$ \\ \hline
        983 & 979 & $1.6\times10^{-6}$ & 982 & $4.4\times10^{-1}$ \\ \hline
        1031 & 1027 & $2.6\times10^{3}$ & 1031 & $7.2\times10^{-3}$ \\ \hline
        1156 & 1160 & $2.9\times10^{7}$ & 1160 & $3.6\times10^{7}$ \\ \hline
        1186 & 1188 & $3.7\times10^{4}$ & 1194 & $1.1\times10^{4}$ \\ \hline
        1209 & 1222 & $1.0\times10^{6}$ & 1211 & $5.3\times10^{-1}$ \\ \hline
        1268 & 1255 & $1.0\times10^{7}$ & 1268 & $2.5\times10^{6}$ \\ \hline
        1279 & 1287 & $1.8\times10^{-2}$ & 1277 & $1.4\times10^{-1}$ \\ \hline
        1309 & 1308 & $4.5\times10^{-1}$ & 1310 & $4.2\times10^{-4}$ \\ \hline
        1329 & 1316 & $8.7\times10^{4}$ & 1328 & $1.4\times10^{5}$ \\ \hline
    \end{tabular}
    \caption[Glucose peaks]{Central frequency of the glucose Raman signals: columns $\nu_{gG},\nu_{gLowBG}$ and $\nu_{gHighBG}$ contain frequencies of correspoding modes in the specified systems in cm$^{-1}$.
    Columns $I_{gHighBG}/I_{gG}$ and $I_{gLowBG}/I_{gG}$ provide relative intensities of the corresponding modes in the B-graphene systems with respect to the intensity of a corresponding mode in the gG systems}
    \label{tab:peaks}
\end{table}

%


\end{document}